\documentstyle[12pt]{article}
\begin{document}
\begin{center}
{\large\bf Planar Two-particle Coulomb Interaction: Classical and Quantum
Aspects}
\end{center}
\vskip 3cm
\noindent
Subir Ghosh,\\
Physics Department,\\
Dinabandhu Andrews College, Garia,\\
Calcutta 700084, India.\\
\vskip 3cm
\noindent
Abstract:\\

The classical and quantum aspects of planar Coulomb interactions have
been studied in detail. In the classical scenario, Action Angle
Variables are introduced to handle relativistic corrections, in the
scheme of time-independent perturbation theory. Complications arising
due to the logarithmic nature of the potential are pointed out.

In the quantum case, harmonic oscillator approximations are considered and
effects of the perturbations on the excited (oscillator) states have been
analysed. 

In both the above cases, the known 3+1-dimensional analysis is carried
through side by side, for a comparison with the 2+1-dimensional (planar)
results.

\newpage
\begin{center}
{\bf I. Introduction}
\end{center}
\vskip .5cm

In recent years, physics of the lower dimensional systems are playing more
and more important roles, not only as toy models for the physical world,
but also as an independent entity. On the one hand, effectively 1+1
and 2+1-dimensional systems have been constructed, while on the other
hand, rich theoretical structures have been unearthed in these low
dimensions.

In the present work, we have studied in some details, the classical and
quantum aspects of planar Coulomb interactions. We have also attempted
to estimate effects of relativistic corrections. The Darwin Lagrangian
\cite {landau},
(and the Hamiltonian), that we have considered has appeared before in
\cite {bg}. Their roles have been further elaborated in 
\cite {subir},
where the Coulomb interaction between anyons \cite {wil}, has been considered.
These studies are done in the spinning particle model of anyon, proposed by
us \cite {sg}.
Unlike the 3+1-dimensional Coulomb problem \cite {gold}, the logarithmic
nature of the Coulomb potential poses computational as well as
conceptual problems. Instead of trying for a full solution of the problem,
we obtain the (degenerate) frequency in the Action Angle Variable (AAV)
formulation \cite {gold}, for a restricted set of orbits. Then we apply the
time independent perturbation theory \cite {gold} in the lowest order
in $\eta=v/c$, ($v$ and $c$ being the velocities of the 
particle and light), for heavy and slowly moving particles, to study
the relativistic corrections. Once again these steps are not
straightforward and we have to make some ansatz which are justified
numerically in a later chapter. We have shown that the perturbation
lifts the degeneracy. However, the relations connecting the original
coordinates to the AAVs are not well defined for
the whole range of variables. Hopefully, this problem might turn out to
be more of a computational nature.

In the quantum case, we have started by performing a harmonic oscillator
approximation about the potential minimum and have compared the 3+1
and 2+1-dimensional results. Interestingly, the binding energy obtained
from classical and semiclassical (or quantum) approaches are not widely
different in the planar case, whereas they differ by orders of magnitude
in the 3+1-dimensional case. The classical and quantum cases differ in
the value of the angular momentum of the particle. In the former, we
obtain it from the particular potential profile we have considered,
whereas in the latter, we simply assume it to be of the order of $\hbar$.
Next we study the effects of perturbation
in the harmonic oscillator excited states \cite {flu}, which contribute
only from the second order in $\eta$. Here we face a conceptual problem
concerning the range of the radial coordinate variable. The correction
depends on the boundary condition, where we fix the zero of the 
logarithmic potential.

As the foreward indicates, it may seem that our analysis has both solved 
and unearthed 
 problems, (at least as far as the study of perturbations are concerned),
 of equal number! This obviously underlines the fact that
we must know the classical and quantum aspects of planar potential
problems more thoroughly. Also it is worthwhile to mention that some of
the troublesome questions are inherently present in 2+1-dimensional
system, which are not a reduction of an original 3+1-dimensional system,
(with a translation symmetry in one direction), as is normally
encountered. 

The paper is organised as follows. In section {\bf II}, we reproduce
the 3+1-dimensional (${1\over r}$) Coulomb problem results in terms of 
AAV. This actually outlines the approach to be pursued for the planar
($ln{r\over{r_0}}$) Coulomb potential, which is carried out in section
{\bf III}. The validity of the ansatz, introduced in section {\bf III},
are checked in section {\bf IV}. Section {\bf V} constitutes a brief
analysis of the time independent perturbation analysis of the relativistic
corrections of the present model, where some problems in the numerical
study have been stressed. Section {\bf V} comprises of the quantum analysis
where again the logarithmic potential raises some non-trivial questions. The
3+1-dimensional analysis is also carried out side by side for completeness. 
\vskip .5cm
\begin{center}
{\bf II. Classical Analysis in 3+1-dimensions}
\end{center}
\vskip .5cm
In this section, we briefly discuss the text book problem \cite {gold}
of 3+1-dimensional Coulomb problem in (AAV), to outline
the approach and for comparison with the 2+1-dimensional case at various
stages.

For the cases where explicit solutions of the time evolution of 
coordinates or the equations of orbits are not required, (or are
not computable), time independent perturbation theory comes to the
rescue. If one can express the original coordinates and momenta in terms of
AAV, the perturbations in the frequencies (of periodic systems) are
computable in a systematic way and with little fuss. However, the catch is
in the former task! As is well known this poses no problem in 3+1-
dimensions, but the analogus problem runs in to severe 
computational trouble in 2+1-dimensions
owing to the logarithmic Coulomb potential.

The Lagrangian and Hamiltonian of two interacting particles of mass $2m$
each and charge $Q$ and $-Q$ are,
\begin{equation}
L={1\over 2}m({\dot r}^2+r^2{\dot{\psi}}^2)+{{Q^2}\over{4\pi\epsilon_0r}},
\label{eql3}
\end{equation}
\begin{equation}
H={{{p_r}^2}\over{2m}}+({{l^2}\over{2mr^2}}-{{Q^2}\over{4\pi\epsilon_0r}})
=KE~+V_{eff},
\label{eqh3}
\end{equation}
where the $r$ and $\psi$ are the relative coordinates in the plane of the 
motion and $m$ is the reduced mass. The conjugate momenta are
$$p_r=m\dot r,~~p_{\psi}=mr^2\dot\psi=l\equiv constant.$$
In Figure (1), 
$~{{4\pi\epsilon_0}\over{Q^2}}V_{eff}=({A\over{r^2}}-{1\over r})$ vs. $r$
is plotted, where $A={{2\pi\epsilon_0l^2}\over{mQ^2}}~metre~=0.01~metre$, 
for later
comparison with 2+1-dimensional results.

The action variables in a general $r,~\theta,~\phi$ coordinate system are
\begin{equation}
J_k=\oint p_kdq_k,~~k=r,~\theta,~\phi,
\label{eqjk}
\end{equation}
the integral being a line integral over complete periods of orbits in
$(q_k,~p_k)$ plane. The generic central force Hamiltonian is
\begin{equation}
H={1\over{2m}}(p_r^2+{{p_{\theta}^2}\over{r^2}}+{{p_{\phi}^2}\over{r^2sin^2
\theta}})+V(r),
\label{eqhh3}
\end{equation}
where $p_k={{\partial L}\over{\partial\dot{q}_k}}$. The total angular
momentum $l$ of (\ref{eqh3}) is related to this notation by $l^2=p_{\theta}^2
+{{p_{\phi}^2}\over{sin^2\theta}}$. Here one obtains,
\begin{equation}
J_\phi=2\pi p_\phi,~~J_\theta=2\pi(l-p_\phi),
\label{eqja3}
\end{equation}
\begin{equation}
J_r=\oint p_rdr=2\int_{r_1}^{r_2}(2mE+{{mQ^2}\over{2\pi\epsilon_0r}}-
{{(J_\theta+J_\phi)^2}\over{4\pi^2r^2}})^{1\over 2}dr,
\label{eqjr3}
\end{equation}
where, $r_1$ and $r_2$ being the two roots of the integrand,
are the turning points of the classical motion and $V(r)$ is the Coulomb
potential. $E$ is the total energy of the system.
This integral can be done directly or via contour method,
with the result,
\begin{equation}
J_r=-(J_\theta +J_\phi)+{{\pi Q^2}\over{4\pi\epsilon_0}}\sqrt{{{2m}\over
{-E}}}.
\label{eqjr}
\end{equation}
Now (\ref{eqjr}) can be inverted to write $E$ in terms of the action
variables only,
\begin{equation}
E=-{{mQ^4}\over{8\epsilon_0^2(J_r+J_\theta +J_\phi)^2}}.
\label{eqe3}
\end{equation}
The degenerate frequency $\nu_k={{\partial E}\over{\partial J_k}}$ is,
\begin{equation}
\nu_r=\nu_\theta=\nu_\phi={{mQ^4}\over{4\epsilon_0^2(J_r+J_\theta+J_\phi)^3}}.
\label{eqf3}
\end{equation}

By performing the indefinite integrals in
\begin{equation}
W=\int^{\phi}p_\phi d\phi+\int^{\theta}p_\theta d\theta+\int^r p_rdr,
\label{eqw3}
\end{equation}
one can see that Hamilton's characteristic function is completely
separable. The angle variables are defined by $w_k={{\partial W}\over
{\partial J_k}}$ and inverting these relations the original coordinates
are expressible in terms of the AAVs. Note that all the integrals are
trivial here. From here the road to time independent perturbation theory is clear
\cite {gold}. The procedure will be elaborated in the main body of our work.

On the other hand, one can 
get the crucial result of (\ref{eqe3}) in an easier
way, albeit in a more restricted setting, by the following argument. The
energy of the {\it circular} orbit is,
\begin{equation}
E_c=H\mid_{p_r=0,~r=r_c}=-{{mQ^4}\over{8\epsilon_0^2(J_\theta +J_\phi)^2}},
\label{eqec3}
\end{equation}
where $r_c$ is the radius of the orbit, $r_c=~2A$, obtained from the
condition $-{{dV_{eff}}\over{dr}}\mid_{r=~r_c}=0$. For circular orbit
we have $p_r=0$ and hence $J_r=0$.

Now if one is assured of the existence of closed orbits, (for example
by invoking Bertrand's theorem \cite {gold}), the frequencies $\nu_k$
must be the same or at most rational multiples of each other. Hence,
at least for the {\it completely degenerate} case, for non-circular
orbits, $J_r$ must enter the energy expression in the same way as
$J_\theta$ and $J_\phi$ in (\ref{eqec3}) and so one recovers the general result
(\ref{eqe3}), by simply replacing  $ (J_{\theta} +J_{\phi})$ by 
$J_r +J_{\theta} +J_{\phi}$ in (\ref{eqec3}).
 Obviously the former rigorous result is much stronger, showing that
there are no non-degenerate orbits in the 3+1 dimensional Coulomb problem.
We will try to apply the latter scheme in the 2+1-dimensional problem under
study.
\vskip .5cm
\begin{center}
{\bf III. Classical Analysis in 2+1-dimensions}
\end{center}
\vskip .5cm
Let us start with the Coulomb Lagrangian and Hamiltonian,
\begin{equation}
{\cal L}={m\over 2}(\dot r^2+r^2\dot\theta^2)-{{Q^2}\over{2\pi\epsilon_0}}
ln{r\over{r_0}},
\label{eql2}
\end{equation}
\begin{equation}
{\cal H}={{{p_r}^2}\over{2m}}+{{Q^2}\over{2\pi\epsilon_0}}({{l^2\pi
\epsilon_0}\over{Q^2mr^2}}+ln{r\over{r_0}})=KE+~V_{eff},
\label{eqh2}
\end{equation}
with $p_\theta=mr^2\dot\theta=~l$, a constant, and $p_r=m\dot r$. 
The Coulomb potential is consistent with the Gauss law,
${\bf\nabla.E}={{\rho}\over{\epsilon_0}}$, where $\rho$ is the charge
density, ${\bf E}$ the electric field and $\epsilon_0$ (or "permittivity")
is just a property of the surrounding vacuum, which is related to the other
property $\mu_0$, 
("permeability"), by the relation $\epsilon_0\mu_0={1\over{c^2}}$.
 Note that now $\epsilon_0$ is such that
${{Q^2}\over{\epsilon_0}}$ has dimension of energy. The Coulomb potential
is attractive for $r<r_0$, $r_0$ being the distance scale where the
potential vanishes.

In Figure (2), we have
shown the potential profile of $V_{eff}={{Q^2}\over{2\pi\epsilon_0}}
({B\over{r^2}}+ln{r\over{r_0}})$ for $B={{l^2\pi\epsilon_0}\over
{mQ^2}}$. The graph is obtained by plotting ${{2\pi\epsilon_0}
\over{Q^2}}V_{eff}$ vs. $r$, with $r_0=1$ and the value of the parameter
$B=10^{-5}~(metre)^2$. Note that $B$ is structurally similar but dimensionally
different from the analogous parameter $A$ in 3+1-dimensions.

The action variables are,
\begin{equation}
J_{\theta}=\oint p_{\theta}d\theta=2\pi l,~~J_r=\oint p_rdr~.
\label{eqj2}
\end{equation}
The straightforward computation of $J_r$ as in (\ref{eqjr3}) is not
possible since the transcendental nature of the integrand forbids explicit
evaluation of the roots (or turning points). A contour integration might
help but we have not been able to perform it. Instead we take recourse to
the qualitative analysis described at the end of section (II).

The energy and radius of the circular orbit are,
\begin{equation}
E_c={{Q^2}\over{2\pi\epsilon_0}}({1\over 2}
+ln[\sqrt{{{\epsilon_0}\over{2\pi m}}}{{J_\theta}\over{Qr_0}}])~
={{Q^2}\over{2\pi\epsilon_0}}({1\over 2}+ln{r\over{r_0}})~,
\label{eqec2}
\end{equation}
\begin{equation}
r_c=\sqrt{{{\epsilon_0}\over{2\pi m}}}{{J_\theta}\over Q}=\sqrt{2B}~.
\label{eqrc2}
\end{equation} 

Bertrand's theorem \cite {gold} assure us that closed orbits are
allowed. Then {\it at least for the degenerate orbits},
where $\nu_r=\nu_{\theta}$, the energy is,
\begin{equation}
E={{Q^2}\over{2\pi\epsilon_0}}({1\over 2}+ln[\sqrt{{{\epsilon_0}\over{2\pi
m}}}{{J_{\theta}+J_r}\over{Qr_0}}])~.
\label{eqe2}
\end{equation}
By inverting this relation we get,
\begin{equation}
J_{\theta}+J_r={{r_0}\over{\sigma}}exp({{2\pi\epsilon_0E}\over{Q^2}})~,
\label{eqj2}
\end{equation}
where $\sigma=\sqrt{{e\epsilon_0}\over{2\pi mQ^2}}$
and $e$ is the exponential number. We will give
numerical estimates of the action integral in (\ref{eqj2}) later
to show the validity of the above ansatz. Hence the single non-trivial
frequency is,
\begin{equation}
\nu_r=\nu_{\theta}={{Q^2}\over{2\pi\epsilon_0}}({1\over{J_r+J_{\theta}}})
={{Q^2}\over{2\pi\epsilon_0J_2}}~,
\label{eqf2}
\end{equation}
where we have defined a new set of action variables, $J_{\theta}\equiv J_1$
and $J_r+J_{\theta}\equiv J_2$. The above is the unperturbed 
degenerate frequency.

The next task is to obtain expressions for the coordinates $r$ and $\theta$
in terms of AAVs. For this we define the Hamilton's characteristic
function,
$$
W=\int^{\theta}p_{\theta}d\theta+\int^rp_rdr$$
\begin{equation}
={{J_1\theta}\over{2\pi}}+\int^r(2mE-{{J_1^2}\over{4\pi^2r^2}}-{{Q^2m}
\over{\pi\epsilon_0}}ln{r\over{r_0}})^{1\over 2}~.
\label{eqw2}
\end{equation}
Note that the arbitrary $r_0$ cancells out in the right hand side, but
we carry it to keep an account of the dimensions and finally take
$r_0=1$metre.

The two angle variables are,
\begin{equation}
w_1={{\partial W}\over{\partial J_1}}={{\theta}\over{2\pi}}-{{J_1}\over
{2\pi^2}}\int {{dr}\over {r^2}}(2mE-{{J_1^2}\over{4\pi^2r^2}}-{{Q^2m}
\over{\pi\epsilon_0}}ln{r\over{r_0}})^{-{1\over 2}}~,
\label{eqw12}
\end{equation}
\begin{equation}
w_2={{\partial W}\over{\partial J_2}}={{Q^2m}\over{\pi\epsilon_0J_2}}
\int dr (2mE-{{J_1^2}\over{4\pi^2r^2}}-{{Q^2m}
\over{\pi\epsilon_0}}ln{r\over{r_0}})^{-{1\over 2}}~
\label{eqw22}
\end{equation}
Once again we are 
stuck with integrals having logarithm inside. Let us rewrite the
dimensionless angle variables in the form,
\begin{equation}
w_1={{\theta}\over{2\pi}}-\sigma J_1\sqrt{{{\pi}\over{2e}}}\int {{dr}\over
{r^2}}(\alpha-ln{r\over{r_0}}-{B\over{r^2}})^{-{1\over 2}}~,
\label{eqww1}
\end{equation}
\begin{equation}
w_2={1\over{2\pi}}{{\sqrt{2e}}\over{\sigma J_2}}\int dr
(\alpha-ln{r\over{r_0}}-{B\over{r^2}})^{-{1\over 2}}~,
\label{eqww2}
\end{equation}
where 
$$\alpha=ln{{\sigma J_2}\over{r_0}},~~ B={{\epsilon_0 J_1^2}\over
{4\pi mQ^2}}={{(\sigma J_1)^2}\over{2e}}.$$
We use the following
arguments to make the substitution,
\begin{equation}
\alpha-ln{r\over{r_0}}-{B\over{r^2}}\equiv (r-r_1)(r_2-r){{2e}\over
{\sigma^2J_2^2}}n^2~,
\label{eqsub}
\end{equation}
where $r_1$ and $r_2$ are the smaller and larger roots and $n$ is an
integer, to be determined later. From the nature of the potential well
in Figure (2), we know that for the energy range $0>E>E_c$, there are 
{\it only two roots}, (or turning points), of the expression of $p_r$
in (\ref{eqw2}). Since in the classical regime, we have to stay inside
the turning points for physical motion, the replacement in (\ref{eqsub})
is justified. Later we will compare the two expressions in (\ref{eqsub})
numerically. The constant factors are picked in such a way that when
(\ref{eqw2}) is inverted, the libration coordinate $r$ becomes a periodic
function of $w_2$. From (\ref{eqww2}) and (\ref{eqsub}), we find,
\begin{equation}
w_2={1\over{2\pi n}}\int {{dr}\over{\sqrt{(r-r_1)(r_2-r)}}}~.
\label{eqwf}
\end{equation}

Next we come to the roots $r_1$ and $r_2$. Indeed, numerical values
of these roots will not suffice, since we require analytic expressions
in terms of AAVs.

Computing the larger root $r_2$ is easier. in the limit $r\rightarrow
\infty$, we get
\begin{equation}
r_2=r_0exp(\alpha)=\sigma J_2~.
\label{eqrl}
\end{equation}
However, estimation of the smaller root $r_1$ is tricky. We use the algebraic
consistency of the above relations, (in the special case of the parameter
value we have chosen from Figure (2)). From (\ref{eqj2}) and (\ref{eqsub}),
we have,
$$J_r=2{{\sqrt e}\over{\sqrt {2}\pi}}\sqrt{{{2\pi mQ^2}\over{e\epsilon_0}}}
\int^{r_2}_{r_1}dr(\alpha-ln{r\over{r_0}}-{B\over{r^2}}^{1\over 2}$$
\begin{equation}
={{2ne}\over{\pi \sigma^2J_2}}\int^{r_2}_{r_1}dr\sqrt{(r-r_1)(r_2-r)}
={{en}\over{4\sigma^2J_2}}(r_2-r_1)^2~.
\label{eqjrf}
\end{equation}
Using (\ref{eqrl}) we simplify (\ref{eqjrf}) to get,
\begin{equation}
J_r={{en}\over 4}(J_2+{{r_1^2}\over{\sigma^2J_2}}-{{2r_1}\over{\sigma}})
={{en}\over 4}(J_r+J_1+{{r_1^2}\over{\sigma^2J_2}}-{{2r_1}\over{\sigma}})~.
\label{eqjrff}
\end{equation}
Hence comparing left and right hand sides, we must have,
$${{en}\over 4}=1,~~r_1^2-2\sigma r_1J_2+\sigma^2J_1J_2=0~.$$
Solving for the nearest integer value of $n$ and for $r_1$ we get,
\begin{equation}
n\approx 2,~~r_1=\sigma J_2\pm\sigma J_2(1-{{J_1}\over{J_2}})^{1\over 2}~.
\label{value}
\end{equation}
When we give the numerical estimates in the next section, it will be seen
that $J_2>J_1$. Since $\sigma J_2=r_2>r_1$, we find
\begin{equation}
r_1\approx \sigma({{J_1}\over 2}+{{J_1^2}\over{8J_2}})~.
\label{eqrs}
\end{equation}
Later we will exihibit values of the roots obtained numerically and
obtained from our approximate analytical expressions, for a comparison.

Taking all this into account, we finally get the angle variable as, (with
$n=2$),
\begin{equation}
w_2={1\over{4\pi}}sin^{-1}{{2r-(r_1+r_2)}\over{r_2-r_1}}~,
\label{w2f}
\end{equation}
$$
w_1={{\theta}\over{2\pi}}-{{\sigma^2J_1J_2}\over{2\sqrt{2}e}}\int
{{dr}\over{r^2\sqrt{(r-r_1)(r_2-r)}}}$$
\begin{equation}
={{\theta}\over{2\pi}}-{{\sigma^2J_1J_2}\over{2\sqrt{2}e}}{{[-r_1r_2+
(r_1+r_2)r-r^2]^{{1\over 2}}}\over{rr_1r_2}}~.
\label{eqw1f}
\end{equation}
By inverting these relations we derive the desired expressions, with
$4\pi w_2=\phi$,
$$r={1\over 2}[(r_2-r_1)sin\phi+(r_1+r_2)]$$
\begin{equation}
={{\sigma J_2}\over 2}[(1-{{\rho}\over 2})sin\phi+(1+{{\rho}\over 2})]~,
\label{eqr}
\end{equation}
\begin{equation}
{{\theta}\over{2\pi}}=w_1+{1\over{6\sqrt{2}e}}\sqrt{{{1-sin\phi}\over{(1+
sin\phi )^3}}}(10+8sin\phi+2sin^3\phi-3\alpha+\alpha sin\phi)~,
\label{eqtheta}
\end{equation}
where $\rho={{J_1}\over{J_2}}$ is a small parameter. Before applying
these relations in perturbation theory, , in the next section we will
demonstrate the validity of our assumptions numerically.
\vskip .5cm
\begin{center}
{\bf IV. Validity of the Ansatz - A Numerical Study}
\end{center}
\vskip .5cm
In the previous section, two crucial assumptions were introduced, in a
rather cavalier fashion: \\
(i) The energy expression of for a circular orbit in (\ref{eqrc2}) was
generalized to any degenerate closed orbit in (\ref{eqe2}).\\
(ii) The expression $\alpha-ln{r\over{r_0}}-{B\over{r^2}}$ was
replaced by a quadratic polynomial in $r$ in (\ref{eqsub}).\\
Below we give the numerical comparison by going through the following
steps.\\
(A): From Figure (2), we obtain a reasonable value of the parameter
$$B={{\epsilon_0J_1^2}\over{4\pi mQ^2}}=10^{-5},~~ r_c=\sqrt {2B}~.$$
For this particular $B$,
$$E_c={{Q^2}\over{2\pi\epsilon_0}}({1\over 2}+ln(\sqrt {2B}))=
{{Q^2}\over{2\pi\epsilon_0}}y~.$$
We have put $r_0=1$ metre.
Hence the range of $E$, such that only two turning points occur, is
$0>E>E_c={{Q^2}\over{2\pi\epsilon_0}}(-4.96)$. 
In Table (1) below, we choose some typical values
for $E$, (where ${{2\pi\epsilon_0E}\over{Q^2}}=y$), 
and the corresponding roots $r_1$ and $r_2$ from Figure (2).\\
\begin{center}
\begin{tabular}{|c|c|c|c|}
\hline
$y$ &  ${r_1}$ &  ${r_2}$ & ${{\sqrt{2e}\over{\pi}}I\mid_n}$\\
\hline
-1 & 0.00133 & 0.3678 & 0.261\\
\hline
-1.5 & 0.0014 & 0.22308 & 0.154\\
\hline
-2 & 0.00149 & 0.13523 & 0.089\\
\hline
-2.5 & 0.00159 & 0.08196 & 0.051\\
\hline
-3 & 0.00172 & 0.04958 & 0.027\\
\hline
-3.5 & 0.0019 & 0.02985 & 0.013\\
\hline
-4 & 0.00216 & 0.01773 & 0.006\\
\hline
-4.5 & 0.00264 & 0.01006 & 0.001\\
\hline
\end{tabular}
\end{center}
(B): First of all, let us check the smallness of $\rho={{J_1}\over{J_2}}$.
For the two extreme values of $E=0$ and $E_c={{Q^2}\over{2\pi\epsilon_0}}
ln(\sqrt{2eB})$, $\rho=\sqrt{2eB} \approx 0.0074$ 
and $\rho_c=1$, (since for circular orbits, $J_1=J_2$.
We have,
$$J_1=\sqrt{{{4\pi B mQ^2}\over{\epsilon_0}}}={{\sqrt{2eB}}\over
{\sigma}}\equiv {x\over{\sigma}}\approx {{0.0074}\over{\sigma}}~,$$
$$J_2=\sqrt{{{2\pi mQ^2}\over{e\epsilon_0}}}exp({{2\pi\epsilon_0 E}
\over{Q^2}})={1\over{\sigma}}exp({{2\pi\epsilon_0E}\over{Q^2}})~.$$
We now compare the above results with our analytical expressions for
$r_1$ and $r_2$ in (\ref{eqrs}) and in (\ref{eqrl}).
Let us take a generic value $E^a$ and the roots $r^a_{1,2}$. Thus we
have $E^a={{Q^2}\over{2\pi\epsilon_0}}y$. We also have 
$$J_1^a={x\over{\sigma}},~~J_2^a={1\over{\sigma}}exp({{2\pi\epsilon_0 E^a}
\over{Q^2}})={1\over{\sigma}}exp(y)~~\rho=xe^{-y}.$$
Hence the roots are,
$$r_1={{\sigma J_1}\over 2}+{{\sigma J_1}\over 8}\alpha=
{x\over 2}+{{x^2 e^{-y}}\over 8}~,$$
$$r_2=\sigma J_2=exp(y)~.$$ 
This are listed in Table (2) below.\\
\begin{center}
\begin{tabular}{|c|c|c|c|}
\hline
$y$ &  ${r_1={x\over 2}(1+{{xe^{-y}}\over 4})}$
&  ${r_2=e^y}$ & ${{{e(r_2-r_1)^2}\over{2\sigma r_2}}}$\\
\hline
-1 & 0.00371 & 0.36788 & 0.49\\
\hline
-1.5 & 0.00372 & 0.22313 & 0.293\\
\hline
-2 & 0.00374 & 0.13534 & 0.174\\
\hline
-2.5 & 0.00377 & 0.08208 & 0.102\\
\hline
-3 & 0.00382 & 0.04979 & 0.058\\
\hline
-3.5 & 0.00391 & 0.0302 & 0.033\\
\hline
-4 & 0.00406 & 0.01832 & 0.015\\
\hline
-4.5 & 0.0043 & 0.01111 & 0.006\\
\hline
\end{tabular}
\end{center}
(C): We here show the comparison between the two forms of $J_r$. One is 
obtained by the following numerical integration, with the limits (obtained
numerically) in Table (1).
$$J_r\mid_n=2\sqrt{{{Q^2m}\over{\pi\epsilon_0}}}\int^{r_2}_{r_1}dr
({{2\pi\epsilon_0E}\over{Q^2}}-{{\pi\epsilon_0J_1^2}\over{4\pi^2mQ^2r^2}}
-ln{r\over{r_0}})^{1\over 2}$$
$$=2\sqrt{{{Q^2m}\over{\pi\epsilon_0}}}\int^{r_2}_{r_1}dr({{2\pi\epsilon_0
E}\over{Q^2}}-{B\over{r^2}}-ln{r\over{r_0}})^{1\over 2}$$
\begin{equation}
=2\sqrt{{{Q^2m}\over{\pi\epsilon_0}}}I\mid_n={{\sqrt{2e}}\over{\pi\sigma}}
I\mid_n~.
\label{eqjrn}
\end{equation}
On the other hand, we have the analytical result from (\ref{eqjrf}),
\begin{equation}
J_r={{en}\over{4\sigma^2J_2}}(r_2-r_1)^2={{e(r_2-r_1)^2}\over{2r_2\sigma}}~.
\label{eqjra}
\end{equation}
This are displayed in Table 2, in order to
 compare between ${{\sqrt{2e}}\over{\pi}}I\mid_n$ and ${{e(r_2
-r_1)^2}\over{2r_2}}$ for different values of $E$ and corresponding
$r_1$ and $r_2$, from the two Tables.\\
(D) In Figure (3), comparative studies of the substitution of the
logarithmic function of $r$ by the quadratic polynomial in $r$ ,
in (\ref{eqsub}), is shown for different values of the energy. 
In fig.(3) we plot the two functions in (\ref{eqsub}) inside the range
$r^a_1$ to $r^a_2$ for four values of $E^a$,
$$f(r)\equiv\alpha-ln{r\over{r_0}}-{B\over{r^2}}=ln{{\sigma J_2}\over r}
-({{\sigma J_1}\over{2er}})^2
=y-ln(r)-{{x^2}\over{4e^2r^2}}~,$$
$$g(r)\equiv{{4e}\over{(\sigma J_2)^2}}(r-r_1)(r_2-r)~
=4ee^{-2y}[r-{x\over 2}(1+{{xe^{-y}}\over 4})][e^y-r].$$
Note
that even though the natures of the curves are somewhat different, (the
respective peaks being in different places), the results are reasonably
satisfactory, since we only use the function inside the action integral
and effects coming from the spatially separated maxima average out.
\vskip .5cm
\begin{center}
{\bf V. Time Independent Perturbation Theory}
\end{center}
\vskip .5cm
In this section, we only reproduce the expressions required for first
order perturbation of the frequency \cite {gold}, where $\eta={v\over c}$
is our small parameter. The perturbation term $H_1$ in $H=H_0+\eta H_1$
is expressed in terms of the unperturbed AAVs,
\begin{equation}
\alpha_1(J_0)={\bar {H_1(w_0,J_0)}}~,
\label{eqph}
\end{equation}
where overbar denotes averaging over a complete period of $w_0$, with
$\nu_0={{\partial H_0}\over{\partial J_0}}$, $w_0=\nu_0t+B_0$, $B_0$
being a time independent 
constant. After the averaging, it is legitimate to replace $J_0$
by $J$, (which is correct to $O(\eta)$), and one derives the frequency
$\nu$, correct upto $O(\eta)$,
\begin{equation}
\nu={{\partial \alpha}\over{\partial J}}=\nu_0+\eta{{\partial \alpha_1}
\over{\partial J}}~.
\label{eqpf}
\end{equation}

The perturbation terms that we consider has been discussed in detail
in \cite {bg} and \cite {subir}.
\begin{equation}
H=H_0+\eta H_1=[{{p^2}\over m}+{{Q^2}\over{2\pi\epsilon_0}}ln{r\over{r_0}}]
+[a(1+ln{r\over{r_0}})+{{jb}\over{mcr}}]~,
\label{eqhp1}
\end{equation}
where $a={{r_ip_i}\over{mcr}}$ and $b={{\epsilon^{ij}r_ip_j}\over{mcr}}$
are dimensionless variables and $j$ is the spin of the 
free particle. In \cite {subir}
it has been shown that in interacting systems of this form, there is a
screening effect on the particle spin.
The higher order terms 
\cite {bg} in $a$ and $b$ are dropped. We replace $p_i$ by $mv_i$ and
note that both $a$ and $b$ are of $O(\eta)$. 
It should be kept in mind that the $a$ dependent
term (and higher order ones present in the original work \cite {bg}) are
conventional ones but the $j$ dependent $b$ terms are the spin effects.
Here our aim is to check the stability of the closed orbits once the 
perturbation is switched on.

It is immedietly apparant from (\ref{eqr}), (\ref{eqtheta}) and (\ref{eqhp1})
that the degeneracy in the frequency is lifted. But we also notice that
due to the necessary averaging to be done with respect to the angle
variable $\phi=4\pi w_2$, the expression for $\theta$ in (\ref{eqtheta})
is not well defined, as the denominator may vanish for some value of the
angle. It should be pointed out that the above mentioned problem crops up
even in the $j$ independent correction terms as well. Hence it is not
clear whether the fault lies in the general perturbation scheme or in
the several approximations that we have introduced.
\vskip .5cm
\begin{center}
{\bf VI. Quantum Mechanics}
\end{center}
\vskip .5cm
We start this section by giving estimates of the binding energy of a
quantum system in 3+1-dimensions. Following the analysis in 
\cite {mes}, the total energy expression to be minimised is
\begin{equation}
E(r=R)={{{\hbar}^2}\over{2mR^2}}-{{Q^2}\over{4\pi\epsilon_0R}}~,
\label{eqquan}
\end{equation}
ehere $R$ is the ground state "radius" of the atom. 
$\hbar$ is a typical value of the angular momentum in
the quantum domain. This gives us
$$R={{4\pi\epsilon_0{\hbar}^2}\over{mQ^2}}\equiv {{8\pi^2{\hbar}^2}
\over e}\bar{\sigma}^2~,$$
$$E(R)=-{{mQ^4}\over{2.(4\pi\epsilon_0\hbar)^2}}\equiv -{{Q^2e}\over
{64\pi^3\epsilon_0{\hbar}^2}}{1\over{\bar{\sigma}^2}}~.$$
This comes to roughly $R\approx 0.529 10^{-10}$ metre and $E\approx
-13.5$ eV. The exactness of the numerical value of $E$ is merely a
coincidence \cite {mes}.

Following this approach in 2+1-dimensions, we minimise,
$$
E(r=R)={{{\hbar}^2}\over{2mR^2}}+{{Q^2}\over{2\pi\epsilon_0}}ln{R\over{r_0}}.
$$
From this we derive,
\begin{equation}
R={{2\pi\hbar}\over {\sqrt{e}}}\sigma,~~E={{Q^2}\over{2\pi\epsilon_0}}
ln{{2\pi\hbar\sigma}\over{r_0}}~.
\label{eqq2}
\end{equation}
The parameter $\sigma$ is expressionwise identical in both cases but
differ in dimension and is the introduced in section (III).

It is amusing to note the following point: 
From Figure (1) and Figure (2), we
can get an order of magnitude idea of the angular momenta involved;
$$l={{\sqrt{eA}}\over{2\pi\bar{\sigma}}}\approx ~2.10^{-30}~Joule-Second,
~~3+1-dim.,$$
\begin{equation}
l={{e\sqrt{2B}}\over{2\pi\sigma}}~.~~2+1-dim.
\label{eqang}
\end{equation}
In estimating the numerical value of $l$ in 3+1-dimension, the particle
is taken to be an electron.
Note that (at least in 3+1-dimension) these values of $l$ are much
larger than $\hbar$, the typical order of quantum angular momentum.
Obviously this $l$ value in 3+1 dimensions will drastically change
the ground state energy from the correct one, making it much smaller.
However, since the angular momentum occures in the logarithm in the
2+1 dimensional energy expression, there is no order of 
magnitude change in the energy value.

One can do a harmonic oscillator approximation about the potential
minima in Figures (1) and (2). We will follow this up in the 2+1
dimensional case with effects of the perturbations introduced in the
previous section.

In 3+1 dimension, we expand the effective potential energy about
$r_c={{\epsilon_0J_1^2}\over{\pi mQ^2}}$,
\begin{equation}
V_{eff}={{Q^2}\over{4\pi\epsilon_0}}(-{1\over{r_c+x}}+{{\epsilon_0J_1^2}
\over{2\pi mQ^2}}{1\over{(r_c+x)^2}})~,
\label{eqhar}
\end{equation}
and identify the quadratic $x$-term with the harmonic oscillator
potential, and obtain the frequency ,
\begin{equation}
\Omega=\sqrt{{Q^2}\over{4\pi\epsilon_0mr_c^2}}={{e\pi Q^2}\over{4\pi
\epsilon_0J_1^3}}{1\over{{\bar\sigma}^2}}={{\pi mQ^4}\over{2\epsilon_0^2
J_1^3}}~.
\label{eqf3}
\end{equation}
Hence the excitation levels above the ground state are
$$E_n=n\hbar\Omega~,~~n=1,~2...~.$$
Identical analysis in 2+1 dimension leads to an expansion of $V_{eff}$
about $r_c=\sqrt{{\epsilon_0}\over{\pi m}}{{J_1}\over Q}$,
\begin{equation}
V_{eff}={{Q^2}\over{2\pi\epsilon_0}}(ln{{r_c+x}\over{r_0}}+{{\epsilon_0
J_1^2}\over{2\pi mQ^2(r_c+x)^2}})~,
\label{eqhar2}
\end{equation}
which produces the frequecny
\begin{equation}
\Omega={e\over{2\sqrt {2}\pi mJ_1\sigma^2}}={{Q^2}\over{\sqrt {2}\epsilon_0
J_1}} ~.
\label{eqe2}
\end{equation}
Note that $\Omega$ is independent of the particle mass.

With the lowest order perturbation terms in (\ref{eqhp1}),
$$H_1={{Q^2}\over{2\pi\epsilon_0}}\eta [(1+ln{r\over{r_0}})cos\theta
+{j\over{mcr}}sin\theta]~,$$
where we have written down the vector products explicitly, we will now
concentrate on
the corrections in the energy, taking the harmonic oscillator excited
states, (above the ground state), as the unperturbed ones. Using the
notation of \cite {flu}, let us consider only the ground state and the
first excited state wave functions.\\
a) Ground state: non-degenerate, $E=\hbar\Omega$, $n=0$, $n_r=0$, $M=0$,
\begin{equation}
u_{0,0}(r,\theta)=\sqrt{{{\lambda}\over{\pi}}}exp(-{{\lambda r^2}\over 2})~,
\label{equ00}
\end{equation}
b) First excited state: doubly degenerate, $E=2\hbar\Omega$, $n=1$,
$n_r=0$, $M=\pm 1$,
\begin{equation}
u_{0,\pm}(r,\theta )={{\lambda}\over{\sqrt{\pi}}}r
~exp(-{{\lambda r^2}\over 2})
exp({\pm i\theta})~.
\label{equ1}
\end{equation}
Here $\lambda={{m\Omega}\over {\hbar}}$ has dimension of $(length)^2$ and
$n$, $n_r$ and $M$ are respectively the total, radial and orbital
angular momentum quantum numbers, with $M$ being an integer.

It is readily seen that $H_1$ above is linear in $cos\theta$ and
$sin\theta$ and hence {\it there are no first order effects in 
any of the states}.

Coming now to second order perturbation theory, we have
\begin{equation}
\Delta_2E_m=\Sigma_{m'}{{\mid<m'\mid V\mid m>\mid ^2}\over{E_m-E_{m'}}}~.
\label{eqp2}
\end{equation}
For the ground state this reduces to
$$
\Delta_2E_0={{\mid<1\mid V\mid 0>\mid ^2}\over{E_0-E_1}}+
{{\mid<-1\mid V\mid 0>\mid ^2}\over{E_0-E_{-1}}}$$
\begin{equation}
={2\over{E_0-E_1}}({{Q^2\eta}\over{4\pi\epsilon_0}})^2[{\cal A}^2+
{{j^2}\over{m^2c^2}}{\cal B}^2]~.
\label{eqgr}
\end{equation}
Here we have used the following relations,
$$<m'\mid cos\theta/sin\theta\mid 0>\equiv \delta_{m',1}\pm\delta_{m',-1}~,$$
$$<\pm1\mid cos\theta\mid 0>={1\over{2\pi}}\int e^{-i(\pm\theta)}cos\theta
d\theta={1\over 2}~,$$
$$<\pm1\mid sin\theta\mid 0>=\pm{i\over 2}~.$$
${\cal A}$, (a dimensionless quantity), and ${\cal B}$, (a quantity of
dimension $(length)^{-1}$), are defined as,
$${\cal A}=<1\mid (1+ln{r\over{r_0}})\mid 0>=2(\lambda)^{{3\over 2}}
\int^{R}_{0}r^2(1+ln{r\over{r_0}})e^{-\lambda r^2}dr~,$$
\begin{equation}
{\cal B}=<1\mid ({1\over r})\mid 0>=2(\lambda)^{{3\over 2}}
\int^{R}_{0}r e^{-\lambda r^2}dr~.
\label{eqAB}
\end{equation}
We are coming to the value of $R$, the upper limit of integration. One
can rewrite the result in (\ref{eqgr}) as,
\begin{equation}
\Delta_2E_0=-{2\over{\hbar\Omega}}({{Q^2\eta}\over{2\pi\epsilon_0}})^2
(\lambda r_0^2)^3(I_1^2+({j\over{mcr_0}})^2I_2^2)~,
\label{eqII}
\end{equation}
where,
\begin{equation}
I_1=\int^{{R\over{r_0}}}_0k^2(1+lnk)exp(-\lambda r_0^2k^2)dk,
\label{eqi1}
\end{equation}
\begin{equation}
I_2=\int^{{R\over{r_0}}}_0k e(-\lambda r_0^2k^2)dk=
{1\over{2\lambda r_0^2}}(1-exp^{-{R\over{r_0}}}~.
\label{eqi2}
\end{equation}
The indefinite integral with the logarithm in (\ref{eqi1}) is not
obtainable in a closed form. We have the following identities \cite {grad},
\begin{equation}
\int^{{R\over{r_0}}}_0~k^2exp(-\lambda r_0^2k^2)dk={1\over 2}
(\lambda r_0^2)^{-{3\over 2}}\gamma({3\over 2},~\lambda R^2),
\label{eqgr1}
\end{equation}
\begin{equation}
\int^{\infty}_0~k^2exp(-\lambda r_0^2k^2)dk={1\over 2}
(\lambda r_0^2)^{-{3\over 2}}\Gamma({3\over 2}),
\label{eqgr2}
\end{equation}
\begin{equation}
\int^{\infty}_0~k^2exp(-\lambda r_0^2k^2)ln(k)dk={{\sqrt\pi}\over 8}
(\lambda r_0^2)^{-{3\over 2}}(2-ln(4\mu)-C),
\label{eqgr3}
\end{equation}
where $\gamma$ and $\Gamma$ are respectively the incomplete Gamma function
and Gamma function and $C$ is the Euler number. Note that in the limit
$R\rightarrow \infty$, $\Delta_2E_0$ in (\ref{eqII}) becomes totally
independent of $r_0$. However, for any finite value of $R$, $r_0$
survives in the $\gamma$ function and in the integral in (\ref{eqgr3})
as well.

The non-trivial issue is the $r_0$ dependence of the final result,
which is coming from the upper limit ${R\over{r_0}}$. If $R$ is
infinity, then the $r_0$-dependence disappears but conceptually
the behaviour of the particles is difficult to visualise: The
attraction (repulsion) between opposite (similar) charges changes to
repulsion (attraction), once the $r_0$-barrier is crossed. It might
seem natural to consider the upper bound of $r$ to be $r_0$, but
this compactification is also not satisfactory.

Generally one encounters logarithmic potentials in physical 
(3 dimensional) space having total infinite amount of charge with
a translation symmetry in one space direction and the questions addressed
here are not always pertinent. But if one is truely in a planar world, with
logarithmic interparticle potential, then these questions can not be
avoided.
\vskip 2cm
Acknowledgement: I am grateful to Professor C. K. Majumdar, Director,
S. N. Bose National Centre of Basic Sciences, for allowing me to use
the Institute facilities.

\newpage
Figure 3\\

The functions,
$$f(r)=y-ln(r)-{{x^2}\over{4e^2r^2}},~~g(r)=4ee^{-2y}
[r-{x\over 2}(1+{{xe^{-y}}\over 4})][e^y-r]$$ 
are plotted for $y=-1,~-2,~-3,~-4$.\\
\vskip 2cm
\begin{center}
(3a)~~~   $y=-1$\\
\end{center}
\vskip 8cm
\begin{center}
(3b)   ~~~$y=-2$\\
\end{center}
\newpage
\begin{center}
(3c)   ~~~$y=-3$\\
\end{center}
\vskip 10cm
\begin{center}
(3d)  ~~~ $y=-4$\\
\end{center}
\newpage
\begin{center}
Figure 1\\
Potential profile in 3+1-dimensions, $A=0.01$.\\
\end{center}
\vskip 10cm
\begin{center}
Figure 2\\
Potential profile in 2+1-dimensions, $B=10^{-5}$.\\
\end{center}
\end{document}